\documentclass[12pt,a4paper,epsf]{article}
\usepackage{amsmath, epsfig}
\begin{document}

\def\NPB#1#2#3{{\it Nucl.\ Phys.}\/ {\bf B#1} (#2) #3}
\def\PLB#1#2#3{{\it Phys.\ Lett.}\/ {\bf B#1} (#2) #3}
\def\PRD#1#2#3{{\it Phys.\ Rev.}\/ {\bf D#1} (#2) #3}
\def\PRL#1#2#3{{\it Phys.\ Rev.\ Lett.}\/ {\bf #1} (#2) #3}
\def\PRT#1#2#3{{\it Phys.\ Rep.}\/ {\bf#1} (#2) #3}
\def\MODA#1#2#3{{\it Mod.\ Phys.\ Lett.}\/ {\bf A#1} (#2) #3}
\def\IJMP#1#2#3{{\it Int.\ J.\ Mod.\ Phys.}\/ {\bf A#1} (#2) #3}
\def\nuvc#1#2#3{{\it Nuovo Cimento}\/ {\bf #1A} (#2) #3}
\def\RPP#1#2#3{{\it Rept.\ Prog.\ Phys.}\/ {\bf #1} (#2) #3}
\def\APJ#1#2#3{{\it Astrophys.\ J.}\/ {\bf #1} (#2) #3}
\def\APP#1#2#3{{\it Astropart.\ Phys.}\/ {\bf #1} (#2) #3}
\def\etal{{\it et al\/}}

\newcommand{\innprod}[2]{\ensuremath{{\langle #1 \mid #2
\rangle}}} 
\newcommand{\bev}{\begin{verbatim}}
\newcommand{\beq}{\begin{equation}}
\newcommand{\beqa}{\begin{eqnarray}}
\newcommand{\beqn}{\begin{eqnarray}}
\newcommand{\eeqn}{\end{eqnarray}}
\newcommand{\eeqa}{\end{eqnarray}}
\newcommand{\eeq}{\end{equation}}
\newcommand{\Eev}{\end{verbatim}}
\newcommand{\bec}{\begin{center}}
\newcommand{\eec}{\end{center}}
\newcommand{\pdif}[2]{\ensuremath{\frac{\partial #1}{\partial #2}}}
\newcommand{\pdiftwo}[2]{\ensuremath{\frac{\partial^2 #1}{\partial
#2^2}}} 
\def\ie{{\it i.e.}}
\def\eg{{\it e.g.}}
\def\half{{\textstyle{1\over 2}}}
\def\nicefrac#1#2{\hbox{${#1\over #2}$}}
\def\third{{\textstyle {1\over3}}}
\def\quarter{{\textstyle {1\over4}}}
\def\m{{\tt -}}
\def\mass{M_{l^+ l^-}}
\def\p{{\tt +}}

\def\slash#1{#1\hskip-6pt/\hskip6pt}
\def\slk{\slash{k}}
\def\GeV{\,{\rm GeV}}
\def\TeV{\,{\rm TeV}}
\def\y{\,{\rm y}}

\def\l{\langle}
\def\r{\rangle}

\begin{titlepage}
\samepage{
\setcounter{page}{1}
\vspace{1.5cm}
\begin{center}
 {\Large \bf Hamiltonian and Potentials in Derivative Pricing Models:\\
Exact Results and Lattice Simulations}

\vspace{1.cm}

 {\large $^1$Belal E. Baaquie, $^2$Claudio Corian\`{o}
  and $^1$Marakani Srikant}

\vspace{.5cm}
{\it $^1$ Physics Department\\
National University of Singapore\\
Singapore, 119260\\
phybeb@nus.edu.sg\\
srikant@srikant.org}

\vspace{.5cm}
{\it $^2$Dipartimento di Fisica\\
 Universita' di Lecce\\
 I.N.F.N. Sezione di Lecce \\
Via Arnesano, 73100 Lecce, Italy\\
claudio.coriano@le.infn.it\\
and \\
Department of Physics, Univ. of Crete, Heraklion, Greece}

\end{center}
\begin{abstract}
The pricing of options, warrants and other derivative securities is one of the 
great success of financial economics. 
These financial products can be modeled and simulated using 
quantum mechanical instruments based on a Hamiltonian formulation.
We show here some applications of these methods for various potentials, 
which we have simulated 
via lattice Langevin and Monte Carlo algorithms,
to the pricing of options. We focus on {\it barrier} 
or {\it path dependent} options, showing in some detail the 
computational strategies involved. 

\end{abstract}
\smallskip}
\end{titlepage}

\section{Introduction} 
Since Black and Scholes' (BS) option pricing model gained an almost immediate acceptance among the professional and academic communities, the trading of derivative securities skyrocketed.
Derivative securities (such as {\it options}) are financial securities whose payoffs 
depends on other underlying securities, and the BS model 
was the first universally accepted modeling of these financial instruments. 

However, in recent years, financial engineeers 
\footnote{Under this professional name  perform actual research 
a large number of physicists and mathematicians, with a remarkable
 number of high energy physicists} 
have created a variety of complex options that are collectively called 
{\it exotic options}. 

The payoffs on these options are considerably more diverse 
than the payoffs on standard BS options or on other straightforward generalizations of them. Some of these financial instruments, widely used in complex portfolios incorporating thousands of these (correlated) elementary instruments, 
may be easier to analize using standard quantum mechanics. 

While we focus on the possibility of applying fundamental theories 
to the real economy, we have to remark that the pattern seems to be 
much wider than expected, since even quantum field theory methods \cite{belal} 
have found their way to the financial world. 

Most of the mathematical methods involved in the analysis 
of financial systems have been based, so far, 
on the simulation of stochastic processes 
by diffusion equations coupled to stochastic sources, i.e. 
stochastic equations of Langevin type. More recently, there has been 
an interest in the analysis of various financial instruments using the path integral formulation. 

Use of a path integral formulation has some advantages. First, it is
in close relation to the lagrangean description of diffusion
processes, second, it opens the way to the use of quantum mechanical
methods, on which we briefly elaborate. 
After a description of the path integral in the Black Scholes model, which we 
study numerically, 
we turn our attention to the analysis of barrier options. 
Barrier options are studied here by simulating an artificial 
quantum mechanical model in which a potential $V(x)$ is added to 
the Black Scholes lagrangean, as first suggested in ref.~\cite{belal}.   
Simulations are carried out using both Langevin and Monte Carlo methods, 
comparing in some limiting cases, where possible, our numerical 
results with analytical ones. 
\section{Langevin Evolution}
In the top down description of theoretical finance, a security $S(t)$
follows a random walk described by a Ito-Weiner process (or Langevin
equation) as
\beq
\frac{d\,S(t)}{S(t)}=\phi d t + \sigma R(t) d t, 
\label{langevin}
\eeq
where R(t) is a Gaussian white noise with zero mean and uncorrelated
values at time t and $t'$ $\langle R(t) R(t')\rangle =\delta(t -t')$.
$\phi$ is the drift term or expected return, while $\sigma$ is a
constant factor multiplying the random source $R(t)$, termed {\it
  volatility}.

As a consequence of Ito calculus, differentials of functions of random
variables, say $f(S,t)$, do not satisfy Leibnitz's rule, and for a
Ito-Weiner process with drift (\ref{langevin}) one easily obtains for
the time derivative of $f(S,t)$
\beq
\frac{d f}{d t}={\partial{f}\over \partial{t}}+ \frac{1}{2}\sigma^2 S^2 
 {\partial^2 f\over \partial{S}^2 }
+ \phi S {\partial{f}\over \partial{S}} + \sigma S 
{\partial{f}\over \partial{S}}R.
\label{deriv}
\eeq
The Black-Scholes model is obtained by removing the randomness of the
stochastic process shown above by introducing a random process
correlated to (\ref{deriv}). This operation, termed {\it hedging},
allows to remove the dependence on the white noise function $R(t)$, by
constructing a {\it portfolio} $\Pi$, whose evolution is given by the
short-term risk free interest rate $r$ 
\beq 
\frac{d\Pi}{dt}= r \Pi
\label{portf}.
\eeq
A possibility is to choose $\Pi=f -{\partial{f}\over \partial{S}}S$. 
This is a portfolio 
in which the investor holds an option $f$ and short sells
\footnote{short selling of the stock should be possible} an amount of the 
underlying security S proportional to ${\partial{f}\over \partial{S}}$. 
A combination of (\ref{deriv}) and (\ref{portf}) yields the 
Black-Scholes equation
\beq
{\partial{f}\over \partial{t}}+ \frac{1}{2}\sigma^2 S^2 
 {\partial^2 f\over \partial{S}^2 }
+ r S {\partial{f}\over \partial{S}}= r f.
\label{bs}
\eeq
There are some assumptions underlying this result. 
We have assumed absence of arbitrage, constant spot rate $r$, 
continuous balance of the portfolio, no transaction costs and 
infinite divisibility of the stock.  

The quantum mechanical version of this equation is obtained by 
a change of variable $S=e^x$, with $x$ a real variable.   
This yields 
\beq
{\partial{f}\over \partial{t}}= H_{BS} f 
\eeq
with an Hamiltonian $H_{BS}$ given by 

\beq
 H_{BS}=- \frac{\sigma^2}{2}\frac{\partial^2}{\partial x^2}
+ \left(\frac{1}{2}\sigma^2 - r\right)\frac{\partial}{\partial x} + r.
\label{bs_hamiltonian}
\eeq
Notice that one can introduce a quantum mechanical formalism and interpret 
the option price as a ket $|f\rangle$ in the basis of $|x\rangle$, 
the underlying security price.  
Using Dirac notation, we can formally reinterpret  
$f(x,t)= \langle x|f(t)\rangle$, as a projection of an abstract quantum state 
$|f(t)\rangle$ on the chosen basis. 

In this notation, the evolution of the option price can be formally written 
as $|f,t\rangle=e^{t H}|f,0\rangle$, for an appropriate Hamiltonian H.  

\section{ Options and Barrier Options }
\subsection{Generalities}
Let the price at time t of a security be $S(t)$. A specific good can
be traded at time t at the price $S(t)$ between a buyer and a seller.
The seller (short position) agrees to sell the goods to the buyer
(long position) at some time $T$ in the future at a price $F(t,T)$ (the
contract price).  Notice that contract prices have a 2-time dependence
(actual time t and maturity time $T$).  Their difference $\tau=T-t$ is
usually called {\it time to maturity}.  Equivalently, the actual price
of the contract is determined by the prevailing actual prices and
interest rates and by the time to maturity.

Entering into a forward contract requires no money, and the value of
the contract for long position holders and strong position holders at
maturity T will be 
\beq (-1)^p\left(S(T) -F(t,T)\right) \eeq 
where $p=0$ for long positions and $p=1$ for short positions.  {\it
  Futures Contracts } are similar, except that the after the contract
is entered, any changes in the market value of the contract are
settled by the parties. Hence, the cashflows occur all the way to
expiry unlike in the case of the forward where only one cashflow
occurs. They are also highly regulated and involve a third party (a
{\it clearing house}). Forward, futures contracts and, as we will see,
{\it options } go under the name of {\it derivative products}, since
their contract price F(t, T) depend on the value of the underlying
security $S(T)$.

In the simplest option, such as a call option, we have seen 
that the payoff function is defined to be the value of the option at maturity 
time $(\tau=0)$. Therefore, the specific path followed 
by the underlying security is not relevant in order to establish 
the price at maturity, except for its final value. 

Barrier options are, instead, path-dependent. This means 
that the payoff is dependent on the realized asset path, 
and certain aspects of the contract are triggered if the asset price, 
from start to end of the contract, becomes too high or too low. 

Barrier options are very popular for various reasons. An investor may have very precise views about the behaviour of a security or he may use them for hedging specific cashflows, to decide to purchase them. In the following, when comparing 
path dependent options to the simplest options, such as standard calls or 
puts, we will refer to the latter as to {\it vanilla} options, using a common financial 
jargon. 

\subsection{ Terminology and Definitions} 
There are some advantages -and natural limitations- in purchasing 
a financial instrument such as a barrier option. 
If the purchaser wants the same payoff typical of a vanilla option, 
but believes that the upward movement of the underlying will 
not be likely, then he may decide to buy an 
{\it up-and-out} call option. The cost of this contract 
will be cheaper than the purchase of a 
corresponding plain vanilla option, but there will be severe limitations 
on the upward movement of the option.

The physical picture of an up-and-out option is that of a 
brownian motion of the underlying asset (x) that is immediately 
killed as soon as the asset hits (from below) the barrier 
B $( x=B)$, which is specified in the contract.

Similarly, a {\it down and out} provision renders the option worthless as soon as 
the asset price hits a barrier B from above. 
The payoffs in the two cases are given by 

\beqa
g_{UO}(x,K) &=&{\it max}(S_T - K)\theta(B - x) \nonumber \\
g_{DO}(x,K) &=&{\it max}(S_T - K)\theta( x - B) 
\label{out}
\eeqa
for a up-and-out (UO) and a down-and-out (DO) option call respectively. 
Here, $\theta()$ denotes the standard step function. A terminology used 
to describe contracts with these features is {\it knocked out options}. 
In contracts of this type it is agreed there will not be any 
payoff if the barrier B is hit. 

Similarly, the market offers contract with additional limitations 
on the allowed variation of the underlying asset. 
For instance, {\it double knock out} 
options have restrictions on the asset variability delimited 
by two barriers ($B_l < B_u$) both from above ($B_u$)  and from below ($B_l$), 
and give zero payoff if any of the two barriers is hit by the asset 
from inception time t to expiry time T. 

{\it Knock in} options are dual, in an obvious sense, to knock out options. 
{\it Knock in} options, in fact, are contracts that pay off as long as the barrier B is hit before expiry. 
If the barrier is hit, then the option is said to have {\it knocked in}, 
otherwise their payoff is null.

Furtherly categorizing these latter types of options, the position 
of the barrier respect to the initial value of the underlying allows 
to distinguish between 
{\it up-and-in} options and {\it down-and-in} options. The payoffs of these 
contracts are given by 

\beqa
g_{UI}(x,K) &=&{\it max}(S_T - K)\theta(x - B ) \nonumber \\
g_{DI}(x,K) &=&{\it max}(S_T - K)\theta(B - x ). 
\label{in}
\eeqa

For definiteness, in the analysis that follows up, we will 
focus our attention to knocked out payoffs of the types described 
in eq.~(\ref{out}).

In knocked out options, single or double, 
killing of the brownian motion is, needless to say, instantaneous, 
and takes place as soon as 
the brownian motion of the asset hits any of the barriers. 

This aspect of the contract is an unpleasent feature since it introduces 
a discontinuity in the dynamics, with attached 
risk management problems both for option buyers and sellers.
Such risks, for instance, are those due to erroneous price movements, or to an instantaneous 
spiky behaviour of an asset, moving upward or downward 
and penetrating a given barrier, which can lead an investor to the loss of all his investment. 
In other unpleasent situations, when large positions of options accumulate 
in the market and are all characterized by the same barrier, 
trading can drive the asset to the barrier, generating massive losses.  

There are various ways by which more conservative and safer 
contracts can be defined, while maintaining some of the features of 
knock out options. 
This is achieved by introducing a finite knock 
out rate, thereby smoothing out the effect of the barrier.  
Our goal is to show how it can be implemented in a self-consistent 
path integral formulation and characterize the pricing 
of these path dependent options.

\section{Quantum Methods in Finance }
To establish a path integral description of a stochastic process we 
need a lagrangean and the corresponding action. This can be easily worked out for the BS model, 
starting from the Hamiltonian given in eq.~(\ref{bs_hamiltonian}). 
We easily gets 
\beq
L_{BS}= -\frac{1}{2 \sigma^2}\left( \frac{d x }{d t } + r -
\frac{1}{2}{\sigma}^2 \right)^2 - r 
\eeq
and the corresponding action, expressed in terms of time to maturity 
$\tau$ 
\beq
S_{BS}=\int_0^\tau L_{BS}\,(t') d\,t'
\eeq
which can be used to define a corresponding path integral for a 
fictitious quantum mechanical process in the variable x, the logarithm 
of the underlying asset

\beq
\langle x_f|e^{-\tau H_{BS}}|x_i\rangle =\Pi_{t_i < t < t_f}
{\int_{-\infty}}^{+ \infty} d\, x(t) e^{S[x]} 
\eeq
with the boundary conditions $x(t_i)=x_i$ and $x(t_f)=x_f$. 
The variable $x=\log(S)$ which identifies the quantum mechanical state of the system will be refered to as to the stock price. The pricing kernel for the 
stock price is given by the 
\beqa
p_{BS}(x,x',\tau)&=&\int DX_{BS}e^{S_{BS}}\nonumber \\
&=&\langle x|e^{-\tau H_{BS}}| x'\rangle\nonumber \\
\eeqa
with 
\beq
\int DX_{BS}=\Pi_{t=0}^{\tau} \int_{-\infty}^{\infty}dx(t). 
\eeq

\subsection{Generalized Potential}

For barrier options it is tempting \cite{belal} to introduce 
a potential $V(x)$ in order to set up a constraint on the 
stochastic process described by the stock price x. 

The corresponding generalized Hamiltonian now reads 
\beq
H_V=-\frac{\sigma^2}{2}\frac{\partial^2}{\partial x^2} + 
\left( \frac{1}{2}\sigma^2 - V(x)\right)\frac{\partial}{\partial x} + V(x).
\eeq
This Hamiltonian is equivalent to a stochastic process, as given in (\ref{disV}), with discounting done by $\exp(-\int dt V(x(t)))$. 

It can be shown \cite{belal} that $H_V$ obeys the martingale condition, and hence can be used for studying processes in finance.

The non-Hermiticity of $H_V$ is of a particularly simple nature, and it can be shown \cite{belal} that for arbitrary $V, H_V$ is equivalent by a similarity transformation to a Hermetian Hamiltonian $H_{\mathrm{Eff}}$ \footnote{Note that for more complex Hamiltonians such as the Merton-Garman $H_{MG}$ finding the equivalent Hermetian Hamiltonian $H_{\mathrm{Eff}}$ is far from obvious} given by 
\begin{eqnarray}
\label{hermh}
H_{\mathrm{Eff}}&=&e^{-s} H_V e^s\\
\mathrm{where} \nonumber\\
H_{\mathrm{Eff}}&=&-\frac{\sigma^2}{2} \frac{\partial^2}{\partial x^2}+\frac{1}{2}\frac{\partial V}{\partial x}+\frac{1}{2\sigma^2}V^2+\frac{1}{2}V+\frac{\sigma^2}{8}\\
\mathrm{and} \nonumber\\
s&=&\frac{1}{2}x-\frac{1}{\sigma^2}\int_0^x dy V(y)
\end{eqnarray}
Note that $H_{\mathrm{Eff}}$ is Hermetian and hence its eigenfunctions form a complete basis; from this it follows that the Hamiltonian $H_V$ can also be diagonalized using the eigenfunctions of $H_{\mathrm{Eff}}$. In particular
\begin{eqnarray}
H_{\mathrm{Eff}}|\phi_n>&=&E_n|\phi_n>\\
\Rightarrow H_V|\psi_n>&=&E_n|\psi_n>\\
\mathrm{where}\nonumber\\
|\psi_n>&=&e^s|\phi_n>\\
<\tilde{\psi}_n|&=&e^{-s}<\phi_n| \ne <\psi_n|
\end{eqnarray}
For the Black-Scholes Hamiltonian $H_{BS}$ we have $V(x)=r$ and hence
\begin{eqnarray}
\label{bseff}
H_{BS}&=& e^s H_{\mathrm{Eff}}e^{-s}\\
&=&e^{\alpha x}\big[-\frac{\sigma^2}{2} \frac{\partial^2}{\partial x^2}+\gamma\big ]e^{-\alpha x}\\
\mathrm{where}\nonumber \\
\gamma&=&\frac{(r+\sigma^2/2)^2}{2\sigma^2}~~~;~~~\alpha=\frac{\sigma^2/2-r}{\sigma^2}
\end{eqnarray}

\section{Path Dependent Options}
We can have very complicated path dependent options since an option is
an arbitrary random variable on the underlying sample space, or in
other words, a completely arbitrary functional of the history of asset
prices. For many but not all kinds of path dependent options, we can
extend the technique of obtaining from the path integral a Hamiltonian
for a quantity related to the option which is path independent and can
therefore be represented as a wave function.  The solution for the
pricing kernel of this quantity then gives us the solution for the path
dependent option. It must be noted that this quantity cannot be a
traded asset as all traded assets evolve with the Black-Sholes
Hamiltonian 
\begin{equation}
  \label{eq:bshamilt}
  \hat{H}_{BS} = -r S \pdif{}{S} - \frac{\sigma^2 S^2}{2} \pdiftwo{}{S}
\end{equation}
according to
\begin{equation}
  \label{eq:hamiltopt}
  \left(-\hat{H} + \pdif{}{t}\right) f = 0
\end{equation}
in the Black-Scholes model. Let us first look at how this is done for
some relatively simple (but more complicated than simple barrier
options) path dependent options.

\subsection{Soft barrier options}
These options have been considered in detail in Linetsky
\cite{Linetsky1}. They are similar to the barrier options considered
above but do not knock out the option completely when the barrier is
hit. Instead, for soft barriers one discounts the final payoff, at some rate, by the exponential of the amount of time spent inside or outside the barrier. For
example, for a down and out barrier is a discounted step option whose barrier is at $B$
and strike price at $K$, the payoff at expiry is
\begin{equation}
  e^{-V \tau_{B_-}} (S_T-K)_+
\end{equation}
where $\tau_{B_-}$ is the time spent below the barrier $B$ and $V$ is
the discounting factor. Considered as a path integral, the current
price of the option is given by 
\begin{equation}
  \int dx' (e^{x'}-K)_+ \int_{x(0) = \ln S}^{x(T) = x'} {\cal D}x e^{S_{BS}} e^{-V \tau_{B_-}} 
\end{equation}
Defining a potential 
\begin{equation}
  \label{eq:potentialstep}
  V(x) = 
  \begin{cases}
    V & x<\ln B\\
    0 & x\ge \ln B\\
  \end{cases}
\end{equation}
we see that that the path integral is equivalent to 
\begin{equation}
  \int dx' (e^{x'}-K)_+ \int_{x(0)=\ln S}^{x(T)=x'} {\cal D}x e^{S} 
\end{equation}
with the action $S$ now being given by 
\begin{equation}
  S = -\int dt L(x, \dot{x}) = - \frac{1}{2\sigma^2} \int dt \left(\left(\dot{x} - r
      -\frac{\sigma^2}{2}\right)^2 + r + V(x)\right)
\end{equation}
In other words, we have just introduced a potential into the
problem. The Lagrangian is now 
\begin{equation}
  L = L_{BS} + V(x)
\end{equation}
and the Hamiltonian now has an extra term $-V(x)$. The new Hamiltonian
is therefore 
\begin{equation}
  \hat{H} = -\frac{\sigma^2}{2} \pdiftwo{}{x} +
  \left(\frac{\sigma^2}{2}-r\right) \pdif{}{x} + r + V(x) 
\end{equation}
Hence, the solution of the step option price in the Black-Scholes
model is equivalent to the solution of the plain vanilla option in a
model with the above Hamiltonian. 

More generally, the pricing of an option whose final payoff is 
\begin{equation}
  e^{-\int V(x(t)) dt} (S_T-K)_+
\end{equation}
is equivalent to the pricing of a plain vanilla call option in a model
where the Hamiltonian is 
\begin{equation}
  \hat{H} = \hat{H}_{BS} + V(x)
\end{equation}
If we can find the eigenvalues and eigenfunctions of the operator
$\hat{H}$, we can write down the pricing kernel using the decomposition
\begin{equation}
  \langle{x}|{e^{-\tau \hat{H}}}|{x'}\rangle = \sum_n \langle{x}|{n}\rangle
    \langle{n}|{e^{-\tau \hat{H}}}|{n}\rangle \langle{n}|{x'}\rangle = \sum_n e^{-\tau
      E_n} \psi_n^*(x') \psi_n(x)
\end{equation}
where the eignevalues of $\hat{H}$ are $E_n$ and the eigenfunctions
corresponding to these eigenvalues are $\psi_n$. When the eignenvalues 
are continuous, the sum becomes an integral
as is the case in the calculations for the single barrier options. In
this case, we can consider the Laplace transform of the pricing kernel
\begin{equation}
  \int_0^\infty d\tau e^{-s\tau} \langle{x}|{e^{-\tau \hat{H}}}|{x'}\rangle 
\end{equation}
which is seen to be the Green's function of the operator $s+\hat{H}$
(this can also be directly seen from the Feynman representation of the pricing kernel). Once we
find the Green's function, we can perform the inverse Laplace
transform to get the pricing kernel and hence the solution to the problem.

To see how this works, let us take the example of the simple barrier
option. We modify the state space using the terms $\alpha$ and $\beta$
so that we only deal with a standard Brownian motion. In that case,
the potential only influences the boundary conditions, so we have to
find the solution to the equation
\begin{equation}
  \left(\frac{1}{2} \frac{d^2}{dx^2} - s\right) G(x,x';s) = -\delta(x-x')
\end{equation}
with the boundary conditions $G(x,x'; s) = 0, x,x' = b = \ln B$ and $\lim_{x
  \rightarrow \infty} G(x,x'; s) = 0$. The Green's functions can be
  easily found using standard methods and the result is given by
\begin{equation}
  \begin{split}
  G(x, x'; s) =& \frac{2\sinh \sqrt{2s}(x-b) e^{-\sqrt{2s}
      (x'-b)}}{\sqrt{2s}} \Theta(x'-x)\\
  &+ \frac{2\sinh \sqrt{2s}(x'-b)
      e^{-\sqrt{2s} (x-b)}}{\sqrt{2s}} \Theta(x-x') 
  \end{split}
\end{equation}
whose inverse Laplace transform is the pricing kernel 
\begin{equation}
  p_{BS}(x,\tau;x')-\frac{1}{\sqrt{2\pi\tau\sigma^2}}e^{-\frac{\tau\beta\sigma^2}{2}-\alpha(x-x')}
  \exp\big[-\frac{1}{2\tau\sigma^2}(x+x'-2B)^2\big]
\label{eq:pkdo}
\end{equation}
where $p_{BS}$ is the Black-Scholes pricing kernel and where the
adjustment for the transformation has been made.

This technique is applied to find a closed form solution for the step
option in Linetsky \cite{Linetsky1}. If we choose the variables such
that the stock price $S=Be^{\sigma x}$ so that the barrier is at $x=0$
and again only deal with the standard Brownian motion, we find that
the Green's function is given by
\begin{equation}
  \label{eq:greenstep}
  G(x, x'; s) = 
  \begin{cases}
    \frac{1}{\sqrt{2s}}\left(e^{\sqrt{2s} |x-x'|} -
      \frac{\sqrt{s+V}-\sqrt{s}}{\sqrt{s+V}+\sqrt{s}}
      e^{\sqrt{2s}(x+x')} \right) & x,x' > 0\\
    \frac{e^{\sqrt{2(s+V)}x - \sqrt{2s}x'}}{\sqrt{2(s+V)} + \sqrt{2s}}
    & x\le 0, x'\ge 0\\
    \frac{e^{\sqrt{2(s+V)}x' - \sqrt{2s}x}}{\sqrt{2(s+V)} + \sqrt{2s}}
    & x\ge 0, x'\le 0\\
    \frac{1}{\sqrt{2(s+V)}}\left(e^{\sqrt{2(s+V)} |x-x'|} -
      \frac{\sqrt{s+V}-\sqrt{s}}{\sqrt{s+V}+\sqrt{s}}
      e^{\sqrt{2(s+V)}(x+x')} \right) & x,x' < 0\\
  \end{cases}
\end{equation}
whose inverse Laplace transform gives us the pricing kernel. 

\subsection{Asian options}
There are several options which cannot be put into a simple form by
discounting alone. One such option which is also fairly popular in the
market is the Asian option which has been considered in detail in the
literature. The Laplace transform of an out of the money Asian option
was found in Geman and Yor \cite{GemanYor}. The payoff of the Asian
option is defined to be
\begin{equation}
  \max\left(0, \frac{1}{T}\int_0^T S(t) dt - K\right)
\end{equation}
We can write the option price as a path integral 
\begin{equation}
  \frac{1}{2\pi} \int dp \int d\nu \int {\cal D}x e^{S_{BS}} e^{ip(\frac{1}{T} \int_0^T dt
    e^{x(t)} - \nu)} (\nu - K)_+ 
\end{equation}
using a standard expression for the Dirac delta function. We see that
we can consider this expression as a plain vanilla call option with a
Lagrangian modified by $-\frac{ip}{T} e^x$. If it were not for the
term $i$, this would be a reducible to a relatively standard
problem. However, since the additional potential term is now complex,
the solution is not so simple. 

One option which is a fairly good approximation for the Asian option
but which is easily solvable is the geometric Asian option. Its final
payoff is defined to be 
\begin{equation}
  \max\left(0, e^{\frac{1}{T} \int_0^T x(t)dt} - K\right)
\end{equation}
To solve this, let us write the Black-Scholes evolution as 
\begin{equation}
  \frac{dx}{dt} = \left(r-\frac{\sigma^2}{2}\right) + \sigma \eta(t) 
\end{equation}
where $\eta(t)$ is white noise. Hence, 
\begin{equation}
  x(t) = x(0) + \left(r - \frac{\sigma^2}{2}\right)t + \int_0^t dt'
  \eta(t') 
\end{equation}
and 
\begin{equation}
  \frac{1}{T}\int_0^T dt x(t) = x(0) + \frac{1}{2}\left(r -
    \frac{\sigma^2}{2}\right) + \frac{\sigma}{T} \int_0^T (T-t) \eta(t) dt
\end{equation}
To find the distribution of the last term, we make use of the
generating function for white noise to give
\begin{equation}
  \frac{1}{2\pi}\int {\cal D}\eta e^{-\frac{1}{2} \int_0^T dt \eta^2(t)}
  e^{ip\left(\frac{\sigma}{T} \int_0^T dt (T-t) \eta(t) - \nu\right)}
\end{equation}
which evaluates to 
\begin{equation}
  \sqrt{\frac{3}{2\pi \sigma^2 T}} e^{-\frac{3 \nu^2}{2\sigma^2 T}}
\end{equation}
In other words, the distribution of $\frac{1}{T}\int_0^T x(t) dt$ is
$N(0, \frac{\sigma^2 T}{3})$. Therefore, the price of the geometric
Asian option is given by 
\begin{equation}
  c = SN(d_1) - Ke^{-r(T-t)}N(d_2)
\end{equation}
where 
\begin{equation}
  d_1 = \frac{\sqrt{3}\left(\ln\left(\frac{S}{K}\right) + \left(\frac{r}{2} +
      \frac{\sigma^2}{12}\right)(T-t)\right)}{\sigma \sqrt{T-t}},\,
  d_2 = \frac{\sqrt{3}\left(\ln\left(\frac{S}{K}\right) + \left(\frac{r}{2} -
      \frac{\sigma^2}{4}\right)(T-t)\right)}{\sigma \sqrt{T-t}} 
\end{equation}

\subsection{Seasoned options}
In the above discussion, we have only discussed how to value path
dependent options where the path dependence starts at the present. In
practice, the problem of solving for the price of path dependent
options after they have been initiated is very important. Such options
are called seasoned options. In many cases, the valuation of seasoned
options proceeds very similarly to that of new options. 

Let us consider a seasoned soft barrier option with discounting by a
potential $V(x)$ at time $t>0$ where the path dependence has started
at zero time. We denote the maturity time be $T$. The value of this
option is given by 
\begin{equation}
  e^{-\int_0^t dt' V(x(t'))} \int {\cal D}x e^{S_{BS}} e^{-\int_t^T dt'
    V(x(t'))} (e^{x(T)}-K)_+ 
\end{equation}
where we should take into account the discounting until time $t$
separately since the history is already known. We see that apart from
this factor, there is no substantial difference between the valuation
of the new and seasoned options. 

For Asian options, we see that we can value the seasoned option
provided we can value the new option since the probability
distribution of the remaining part of the average will determine the
option price at the time when combined with information about the
contribution to the average of the revealed historical price. Hence,
we see that we can price seasoned Asian options if we can price new
Asian options. 
\section{Solving the double knock out barrier option}

A double barrier option is an option whose value reduces to zero
whenever the price of the underlying instrument hits the barriers
which we denote by $e^a$ and $e^b$. Hence, the price of a double knock
out barrier European call option expiring at time $T$ and with strike
price $K$ at time $t_0$ provided it has not already been knocked out
will be given by
\begin{equation}
  \label{eq:double_bar_def}
  e^{-r(T-t_0)}E_t[(e^{x(T)}-K)_+]{\mathbf 1}_{a<x(t)<b, t_0<t'<T}
\end{equation}
where ${\mathbf 1}$ stands for the indicator function. It is
sufficient to solve for the probability distribution of $x(T)$ for
those paths which do not go outside the barriers (in other words, the
pricing kernel). 

Written as a path integral, the formula is 
\begin{equation}
  \label{eq:pathint_doub_bar}
  e^{-r(T-t_0)} \int {\cal D}x \Theta (x(t)-a) \Theta (b - x(t))
  e^{S_{BS}(x(t))} (e^{x(T)}-K)_+ 
\end{equation}
where $S_{BS}$ is the Black-Scholes action 
\begin{equation}
  \label{eq:bsaction}
  S_{BS} = -\frac{1}{2\sigma^2}\int dt (\dot{x} + r -
  \frac{\sigma^2}{2})^2 
\end{equation}

While the step functions look complicated in the path integral, they
can be seen to be having the effect of an infinite potential barrier
since they effectively prohibit the path from entering the forbidden
region outside the barriers. Hence, the problem might be better solved
using the Hamiltonian and this is indeed the case. 

In the Schr\"odinger formulation, the above problem is to find the
pricing kernel for a system with the Hamiltonian 
\begin{equation}
  \label{eq:hamiltonian}
  \hat {H} = \hat {H}_{BS} + V(x) 
\end{equation}
where the Black-Scholes Hamiltonian is given by 
\begin{equation}
  \label{eq:hbs}
  \hat {H}_{BS} = -\frac{\sigma^2}{2} \frac{\partial}{\partial x^2} +
  (\frac{\sigma^2}{2} - r) \frac{\partial}{\partial x}
\end{equation}
and the potential $V(x)$ is given by
\begin{equation}
  \label{eq:potential}
  V(x) = 
  \begin{cases}
    \infty & x<a\\
    0 & a<x<b\\
    \infty & x>b
  \end{cases}
\end{equation}

This is very similar to the well known problem of a particle in an
infinite potential well except that the Hamiltonian has an extra term
involving $\frac{\partial}{\partial x}$ which makes it non-Hermitian.

This problem can be solved by transforming the underlying wave
functions. By making the transformation $\innprod{x}{\phi} =
e^{-\alpha (x-a)} \innprod{x}{\psi}$ and $\innprod{\phi}{x} =
e^{\alpha (x-a)} \innprod{\psi}{x}$, where $|\phi \rangle$ are the
vectors in the new (Hilbert) space, $|\psi \rangle$ and $\langle
\tilde{\psi}|$ are the original vectors and their duals respectively
and $\alpha = \frac{\sigma^2/2-r}{\sigma^2}$. In this new space, the
Black-Scholes Hamiltonian takes the simple Hermitian form
$-\frac{\sigma^2}{2} \frac{\partial^2}{\partial x^2}$.

The problem is now identical to that of a quantum mechanical particle
of mass $\frac{1}{\sigma^2}$ (in units where $\hbar=1$) in an infinite
potential well. As is well known in this case, the allowed momenta are
$p_n = \frac{n\pi}{b-a}$. The eigenfunctions are hence given by
\begin{eqnarray}
  \label{eq:eigenf}
  \innprod{x}{\psi_n} &= e^{\alpha(x-a)} \innprod{x}{\phi_n} &=
  \sqrt{\frac{2}{b-a}}ie^{\alpha (x-a)} \sin{p_n (x-a)}\\ 
  \innprod{\tilde{\psi}_n}{x} &= e^{-\alpha (x-a)}
  \innprod{\phi_n}{x} &= -\sqrt{\frac{2}{b-a}}ie^{-\alpha (x-a)} \sin{p_n(x-a)}
\end{eqnarray}
where $\innprod{x}{\phi_n}$ are the eigenfunctions of the quantum
mechanical particle in an infinite potential well. 

The eigenfunctions are orthonormal 
and form a complete basis since 
\begin{equation}
  \begin{split}
    \label{eq:completeness}
    &\sum_{n=1}^\infty  \innprod{x}{\psi_n} \innprod{\tilde{\psi}_n}{x'} =
    \frac{2}{b-a} e^{\alpha (x-x')} \sum_{n=1}^\infty \sin{p_n (x-a)} \sin{p_n
      (x'-a)} \\
    &= \frac{1}{2(b-a)} e^{\alpha (x-x')} \sum_{n=-\infty}^\infty
    \left(\exp{\frac{in\pi}{b-a} (x-x')} - \exp{\frac{in\pi}{b-a}
      (x+x'-2a)}\right)\\
    &= \frac{\pi}{b-a} e^{\alpha (x-x')} \left(\delta\left(\frac{\pi (x-x')}{b-a}\right) -
    \delta\left(\frac{\pi (x+x'-2a)}{b-a}\right)\right)\\
    &= \delta(x-x')
  \end{split}
\end{equation}
since $a<x<b$ and $a<x'<b$. 

The pricing kernel is hence given by 
\begin{equation}
  \label{eq:pricing kernel}
  \begin{split}
    \langle{x} |e^{-\tau \hat{H}}|{x'}\rangle =
& \sum_{n=1}^\infty \sum_{n'=1}^\infty
     \langle{x}|{\psi_n}\rangle\langle{\tilde{\psi}_n}|{e^{-\tau \hat{H}}}|{\psi_{n'}}\rangle
     \langle{\tilde{\psi}_{n'}}|{x'}\rangle\\
     =&\sum_{n=1}^\infty \innprod{x}{\psi_n} \innprod{\tilde{\psi}_n}{x'} e^{-\tau E_n}\\
     =&\frac{1}{2(b-a)}\exp{\left(-\frac{\tau \sigma^2 \beta}{2} +
         \alpha (x-x')\right)}\\ 
     &\sum_{n=-\infty}^\infty \exp{\left(-\frac{\tau
           \sigma^2 p_n^2}{2}\right)} (e^{ip_n(x-x')} - e^{ip_n (x+x'-2a)})\\
     =& \frac{1}{2(b-a)} \exp{\left(-\frac{\tau \sigma^2 \beta}{2} +
         \alpha (x-x')\right)} \sum_{n=-\infty}^\infty \int dy \delta
     (y-n) \exp{\left(-\frac{y^2 \pi^2 \tau \sigma^2}{2(b-a)^2}\right)}\\
     & \left(\exp{\frac{iy\pi (x-x')}{b-a}} - \exp{\frac{iy\pi
           (x+x'-2a)}{b-a}} \right)\\
     =& \sqrt{\frac{1}{2\pi \tau \sigma^2}}
     \exp{\left(-\frac{\tau \sigma^2 \beta}{2} + \alpha(x-x')\right)}\\
    &\sum_{n=-\infty}^\infty \left( \exp{-\frac{(x-x' +
          2n(b-a))^2}{2\tau \sigma^2}} - \exp{-\frac{(x + x' - 2a -
          2n(b-a))^2}{2\tau \sigma^2}}\right)
  \end{split}
\end{equation}
where 
\begin{equation}
  \label{eq:beta}
  \beta = \frac{(\sigma^2/2 + r)^2}{\sigma^4}
\end{equation}
and the identity 
\begin{equation}
  \label{eq:ident}
  \delta (y-n) = \sum_{n=-\infty}^\infty e^{2\pi iny}
\end{equation}
has been used.

Hence, we see that the pricing kernel (apart from the drift terms) is
given by an infinite sum of Gaussians. To check its reasonableness, we
check the value in the limits $b\rightarrow \infty$ and $a \rightarrow
-\infty$. In the former case, only the $n=0$ term contributs and in the
latter, only the $n=0$ and $n=1$ terms contribute. It is easy to see
that, in both cases, the result reduces to the solution for the single
knockout barrier pricing kernel. When both limits are simultaneously
active, only the first term in the $n=0$ term exists and it is easily
seen that gives rise to the well known Black-Scholes pricing kernel. 

We can now evaluate the price of a double barrier European call option
using the pricing kernel from (\ref{eq:pricing kernel}). The result is seen to
be 
\begin{equation}
  \label{eq:option}
  \begin{split}
    f = &\sum_{n=-\infty}^\infty \left(e^{-2n\alpha(b-a)}
    \left(e^{2n(b-a)}SN(d_{n1})- Ke^{-r\tau} N(d_{n2})\right)\right.\\
    &- \left.S^{2\alpha} e^{-2\alpha(n(b-a)-a)}
    \left(e^{2n(b-a)} \frac{e^{2a}}{S} N(d_{n3}) - Ke^{-r\tau}
      N(d_{n4})\right)\right)
  \end{split}
\end{equation}
where
\begin{eqnarray}
  \label{eq:dns}
  d_{n1} &=& \frac{\ln(\frac{S}{K}) + 2n(b-a) + \tau
    \left(r+\frac{\sigma^2}{2}\right)}{\sigma \sqrt{\tau}}\\
  d_{n2} &=& \frac{\ln(\frac{S}{K}) + 2n(b-a) + \tau \left(r -
      \frac{\sigma^2}{2}\right)}{\sigma \sqrt{\tau}} =
  d_{n1} - \sigma \sqrt{\tau}\\
  d_{n3} &=& \frac{\ln(\frac{e^{2a}}{SK}) + 2n(b-a) +
    \tau\left(r+\frac{\sigma^2}{2}\right)}{\sigma \sqrt{\tau}}\\
  d_{n4} &=& \frac{\ln(\frac{e^{2a}}{SK}) + 2n(b-a) +
    \tau\left(r-\frac{\sigma^2}{2}\right)}{\sigma \sqrt{\tau}} =
  d_{n3} - \sigma \sqrt{\tau}
\end{eqnarray}

\section{Monte Carlo Simulations}
The pricing kernel is the fundamental quantity to compute using the functional 
integral. Related attempts can be found in the literature \cite{NM}.
We assume a discretization of the time to maturity $\tau$ in intervals 
$\epsilon=\tau/N$, with N an arbitrary (large) integer.  

For instance,  for the Black Scholes model one gets the action 
\beq
S_{BS}= \epsilon \sum_{i=1}^N L_{BS}(i)
\eeq
with
\beq
L_{BS}(i)=-\frac{1}{2 \sigma^2}\left( \frac{x_i - x_{i-1}}{\epsilon}
 + r - \frac{\sigma^2}{2}\right)^2
\eeq
where we have introduced discretized positions $(x_i)$ 
for the variable 

For this purposes, we have used a standard Metropolis algorithm. 
If thermalization is slow, it is possible to resort to use 
sequentially Metropolis updates and cluster updates. 
The latter is an update for the embedded Ising dynamics in the lattice 
variables $x_i/|x_i|$ 
(Swendsen-Wang, Wolff), and is included in for a faster generation of the 
thermalized paths of the stock price $x(t)$.
  
For processes involving a stochastic volatility $(y=\log(V))$ 
the expression of the path integral is more complicated and can be found 
in \cite{BM}. From now on we will just consider the case of a 
constant volatility.
 
If we denote by $g(x,K)$ the payoff function, with a strike price $K$, 
in this case the value of the option (its price) is given by the Feynman-Kac formula 
\beq
f(t,x)=\int_{-\infty}^{\infty}d\,x'\langle x|e^{-(T-t)H_{BS}}|x'\rangle g(x',K)
.
\label{FK}
\eeq
In actual simulations, it is convenient to compute directly 
the option price rather than the propagator itself. 
The simulation is done by taking the initial point x fixed, and letting the 
final point evolve according to its quantum dynamics. In this way a path 
$(x,x')$ is generated. After the first thermalization, $x'$ is allowed to 
undergo quantum fluctuations, at fixed x. Each $x'$ is then 
convoluted with the payoff function and an average is performed. Finally, this 
procedure is repeated for several x values, so to obtaint the option price 
at time to maturity $\tau$.
\begin{figure}[th]
\centerline{\includegraphics[angle=-90,width=.8\textwidth]{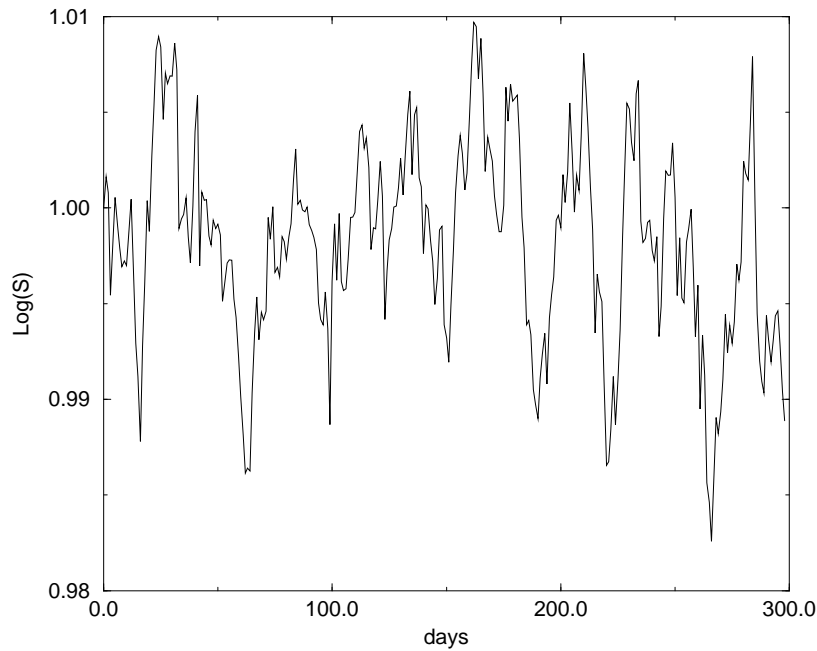}}
\caption{An example of thermalized path obtained from the simulation of the path integral (Black-Scholes) with r=0.05 and $\sigma=0.12$ }
\end{figure}
\begin{figure}[th]
\centerline{\includegraphics[angle=-90,width=.8\textwidth]{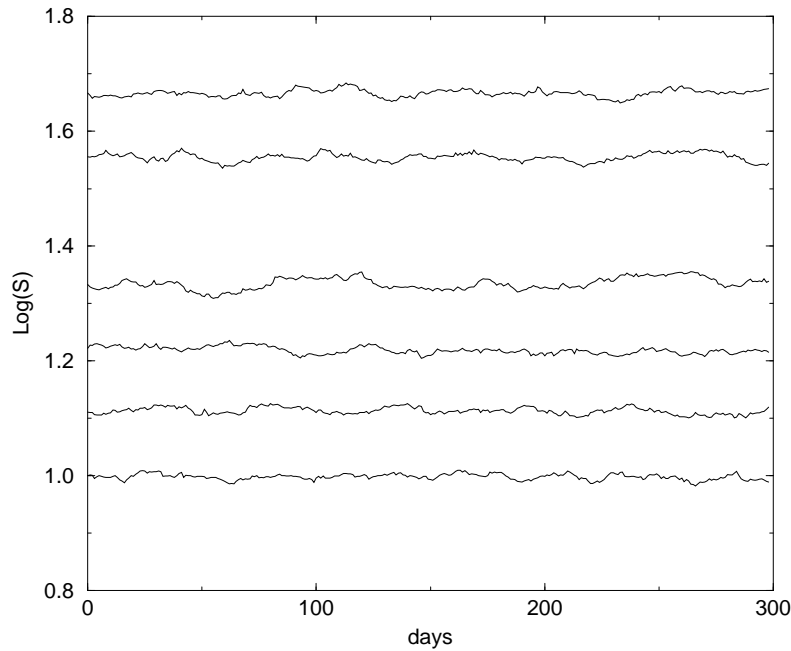}}
\caption{Several thermalized paths for (Black-Scholes) with r=0.05 and 
$\sigma=0.12$ }
\end{figure}

\begin{figure}[th]
  \begin{center}
    \epsfig{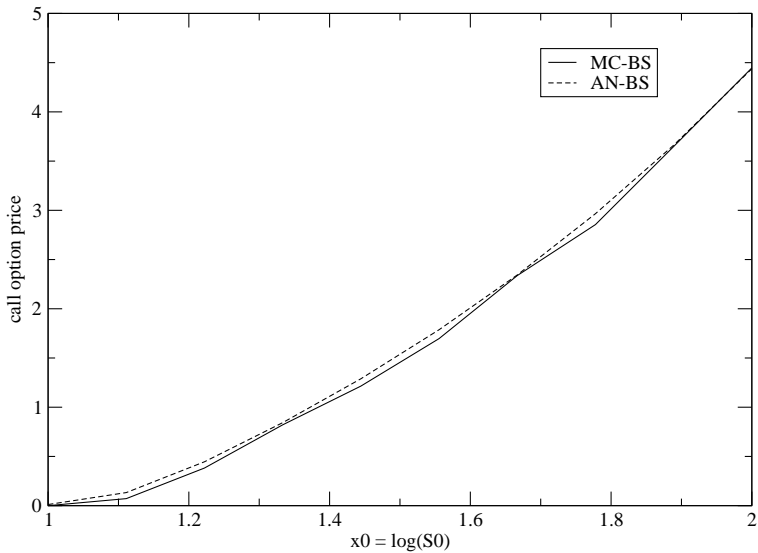}
    \caption{Call option price for strike price 3 versus the logarithm of the initial  value of the stock $x_0=\log(S_0)$. 
      the parameters are fixed as in figs~1. Shown is the analytical result 
      vs the monte carlo result, with a low resolution of 10,000
      configurations}
  \end{center}
\end{figure}
Figs. 1, 2 and 3, 
illustrate some simple results obtained by the monte carlo method. 
For illustrative purposes, we show the behaviour of the 
Black-Scholes model. Fig.~1 shows a typical thermalized path, generated from a given initial value x (at current time $t=\tau$) assuming a maturity of 300 days, while in Fig.~2 we have plotted several path for different starting 
values x of the stock at current time $\tau$. We have chosen an interest 
rate $r=0.05$ and a 12 percent volatility $\sigma$.
Finally, in Fig.~3 we compare the analytical and the numerical 
evaluation of the Black-Scholes option price with a low resolution for 
(\ref{FK}), in order 
to separate the two curves, which otherwise would 
overlap completely, in order to illustrate the convergence of the 
Metropolis algorithm. 

\section{Langevin Calculation of Option Prices with Potential: 
Numerical Methods}

As is well known from quantum mechanics, the path integral 
\begin{equation}
  \label{eq:qm}
  \int {\cal D}x e^{\frac{i}{\hbar}\int L(x,\dot{x}) dt}
\end{equation}
with the Lagrangian $L$ given by 
\begin{equation}
  \label{eq:lagrangian}
  \frac{1}{2} m\dot{x}^2 - V(x) 
\end{equation}
is the pricing kernel for a one-dimensional system of a particle of mass
$m$ moving in a potential $V(x)$. This is very similar to the path
integral for standard Brownian motion $W(t)$ given by 
\begin{equation}
  \int {\cal D}W e^{-\frac{1}{2}\int \dot{W}^2(t) dt}
\end{equation}
The path integral for a more general  stochastic process described by
the stochastic differential equation (in physics, the equation is
usually written out with everything divided by $dt$ and with
$dW(t)/dt$ replaced by $\eta$ which represents white noise and the
equation is then called the Langevin equation) 
\begin{equation}
  \label{eq:gensde}
  dx(t) = a(x)dt + \sigma(x) dW(t) 
\end{equation}
The Lagrangian can be found by solving the above for $\dot{W}$. One
easily obtains the path integral
\begin{equation}
  \int {\cal D}x e^{-\int dt \frac{1}{2\sigma^2(x(t))}  (\dot{x} - a(x(t)))^2}  
\end{equation}
When we perform a Wick rotation for the action in (\ref{eq:qm}), we
get the Lagrangian for Brownian motion if $V(x)=0$. Hence, a free
quantum mechanical particle can in some sense be considered to be
undergoing Brownian motion and can be modelled by the stochastic
differential equation 
\begin{equation}
  \label{eq:qmsde}
  dx(t) = \sqrt{\frac{\hbar}{m}}dW(t)
\end{equation}

The path integral for the Euclidean action can be numerically
simulated using the Metropolis algorithm. Another method is to use the
analogy and directly integrate the stochastic differential equation
(\ref{eq:qmsde}). The latter is more efficient as one does not have
the problem of correlation between succesive configurations which
reduces the accuracy of Monte Carlo calculations. There does seem to
be a problem in that potentials cannot be included. However, this can
be handled by including a killing term in the stochastic differential
equation whose connection with the potential becomes clear when we
consider the Hamiltonian. The integration of stochastic differential
equations can also give efficient numerical calculations in quantum
field theory, especially for gauge invariant theories where gauge
fixing is trivial in this framework. This is dealt with in great
detail in Namiki \cite{Namiki}.

In the case of stock option pricing, we are usually dealing with the
Black-Scholes stochastic differential equation 
\begin{equation}
  dS(t) = rS(t)dt + \sigma S(t) dW(t)
\end{equation}
whose Hamiltonian is given by 
\begin{equation}
  \frac{1}{2} \sigma^2 S^2 \pdiftwo{}{S} + rS \pdif{}{S} 
\end{equation}
and Lagrangian by 
\begin{equation}
  -\frac{1}{2\sigma^2 S^2} \left(\dot{S} - rS\right)^2 
\end{equation}
This becomes much simpler when we transform the stochastic
differential equation to the variable $x = \ln S$ using It\^o's
lemma to give 
\begin{equation}
  dx(t) = \left(r - \frac{\sigma^2}{2}\right) dt + \sigma dW(t)
\end{equation}
with the Hamiltonian 
\begin{equation}
  \frac{1}{2} \sigma^2 \pdiftwo{}{x} + \left(r -
    \frac{\sigma^2}{2}\right) \pdif{}{x}
\end{equation}
and much simpler Lagrangian
\begin{equation}
  -\frac{1}{2\sigma^2} \left(\dot{x} - r + \frac{\sigma^2}{2}\right)^2
\end{equation}
We will consider a somewhat more general stochastic process for the
Black-Scholes stochastic differential equation given by 
\begin{equation}
\label{disV}
  dx(t) = \left(V(x) - \frac{\sigma^2}{2}\right)dt + \sigma dW(t)
\end{equation}
which represents a situation where $r$ is given by $V(x)$. We can recover the Black-Scholes result by setting $V(x)=r$ and a general potential $V(x)$ is interesting as a mathematical exercise. 

To accomodate the discounting of all assets by the money market
account, we have to include a killing term in the stochastic
differential equation so that all expectations are discounted by
$\exp(-\int dt V(x(t)))$. This changes the Hamiltonian to 
\begin{equation}
  \frac{1}{2} \sigma^2 \pdiftwo{}{x} + \left(V(x) -
    \frac{\sigma^2}{2}\right) \pdif{}{x} - V(x)  
\end{equation}
The reason the Hamiltonian above differs by an overall sign from the Black-Scholes Hamiltonian is because in pricing of options, one is considering the {\it backward} Fokker-Planck equation that results from the stochastic Langevin equation. 

The Hamiltonian above can be simulated by numerically integrating the Black-Scholes
stochastic differential equation with $r$ replaced by $V(x)$ and
putting in the discount factor explicitly when calculating
expectations.

Alternatively, general barrier options can be considered by keeping
the Black-Scholes process for $x$ but introducing an extra killing
term $V(x)$. In that case, we get the Hamiltonian 
\begin{equation}
  \label{eq:hamilt1}
  \hat{H} = -\frac{\sigma^2}{2}\frac{\partial^2 }{\partial x^2} +
  (\frac{\sigma^2}{2}-r)\frac{\partial}{\partial x} + r + V(x)
\end{equation}
where the killing term $r$ is included because even the plain vanilla
option price must be discounted by the money market account to get its
current price. This can also be simulated by numerically integrating
the Black-Scholes stochastic differential equation and putting in the
discount factor explicitly when calculating expectations.

In general, therefore, option pricing with the Hamiltonian
\begin{equation}
  \label{eq:hamilt2}
  \hat{H} = -\frac{\sigma^2}{2}\frac{\partial^2 }{\partial x^2} +
  ( \frac{\sigma^2}{2}-r)\frac{\partial}{\partial x} + r + V(x)
\end{equation}
is equivalent to solving the It\^o stochastic differential equation
\begin{equation}
  \label{eq:sde1}
  dx = \left(r-\frac{\sigma^2}{2}\right)dt + \sigma dW(t)
\end{equation}
where $W(t)$ is a standard Wiener process and then taking all
expectations discounted by $\exp(-\int (r+V(x(t))) dt)$
\cite{Oksendal}. If we are using the Hamiltonian
\begin{equation}
  \label{eq:hamilt3}
  \hat{H} = -\frac{\sigma^2}{2}\frac{\partial^2 }{\partial x^2} +
  ( \frac{\sigma^2}{2}-V(x))\frac{\partial}{\partial x} + V(x)  
\end{equation}
this becomes equivalent to solving the stochastic differential
equation 
\begin{equation}
  \label{eq:sde2}
  dx = (V(x)-\frac{\sigma^2}{2})dt + \sigma dW(t)
\end{equation}
with discounting now being done by the factor $\exp(-\int
V(x(t))dt)$.

This can be done in a straightforward manner numerically. The
stochastic differential equation can be solved by the Euler method to
give sample paths $x(t_i), 0\le i\le N$ where $t_0 = 0$,
$t_i=\frac{iT}{N} = i\epsilon$ and the discounting of the final value
of the option can be done with the factor given above. More
explicitly, the simulation is done by numerically integrating the
It\^o stochastic differential equation (\ref{eq:sde1}) using
\begin{equation}
  \label{eq:numhamilt1}
  x_{i+1} = x_i + (r - \frac{\sigma^2}{2}) (\epsilon) +
  \sigma \sqrt{\epsilon} Z
\end{equation}
or (\ref{eq:sde2}) using 
\begin{equation}
  \label{eq:numhamilt2}
  x_{i+1} = x_i + \left(V(x_i) - \frac{\sigma^2}{2}\right) (\epsilon) +
  \sigma \sqrt{\epsilon} Z  
\end{equation}
where $x_i$ is $x(t_i)$ and $Z$ are normally distributed random
numbers with unit variance. Such random numbers can be obtained from
the usual uniformly distributed random numbers in the range $[0,1)$ by
the transformation
\begin{eqnarray}
  \zeta_1 &= \sqrt{-2 \ln \xi_1} \cos 2\pi \xi_2\\
  \zeta_2 &= \sqrt{-2 \ln \xi_1} \sin 2\pi \xi_2
\end{eqnarray}
where $\xi_1$ and $\xi_2$ are distributed uniformly on $[0, 1)$ and
$\zeta_1$ and $\zeta_2$ are now normally distributed with unit
variance. 

The modified call option price desribed by (\ref{eq:hamilt1}) is then
given by
\begin{equation}
  E\left[\exp\left(-rT - \epsilon \sum_{i=0}^{N-1} V(x_i)\right) \left(e^{x_N}-K\right)_+\right]
\end{equation}
where the $x$ are generated using (\ref{eq:numhamilt1}). The modified call
option prices described by (\ref{eq:hamilt2}) are given by
\begin{equation}
  E\left[\exp\left(-\epsilon \sum_{i=0}^{N-1} V(x_i)\right) \left(e^{x_N}-K\right)_+\right]
\end{equation}
where the $x$ are generated using (\ref{eq:numhamilt2}). 

\section{Results of Numerical Simulations}
The initial stock price $S_0$ is assumed to be 100. We make use of the
variable $x=\ln S$ as explained above and we set $x_0 = \ln S_0$.

There are two ways we use the potential as explained in the previous
section. In the first case, the killing term (coefficient of constant)
in the Hamiltonian is $r+V(x)$ while the drift term, the coefficient
of $\pdif{}{x}$, is still $r-\frac{\sigma^2}{2}$. This corresponds to
a constant interest rate $r$ and where the option payoff at expiry is
defined in a path dependent way as
\begin{equation}
  g[S] = \left(\exp\left(-\int_0^T V(S(t)) dt\right) S(T)-K\right)_+
\end{equation}
where $(y)_+$ stands for $\max(0,y)$. In this case, the potential $V$
and $r$ are separated as they have different interpretations. 

In the second case, the interest rate is assumed to be given as a
function of the underlying security price and is equal to $V(x)$. In
this case, the Hamiltonian has $V$ in both the drift and killing
(coefficient of constant) terms. The option payoff $g(S)$ is then {\em
  path-independent} and is the usual call option payoff $\max(0,
S-K)$. 

\begin{figure}[h]
  \centering
  \epsfig{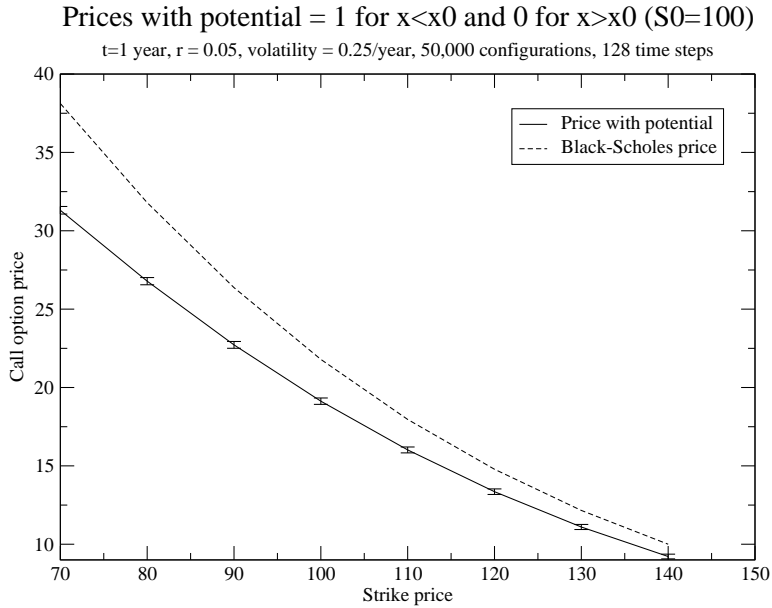}
  \caption{Plot of call option price against strike price for a purely
    discounting potential of form (\ref{eq:one}) and for the Black-Scholes case.}
  \label{fig:one}
\end{figure}

\begin{figure}[h]
  \centering
  \epsfig{file=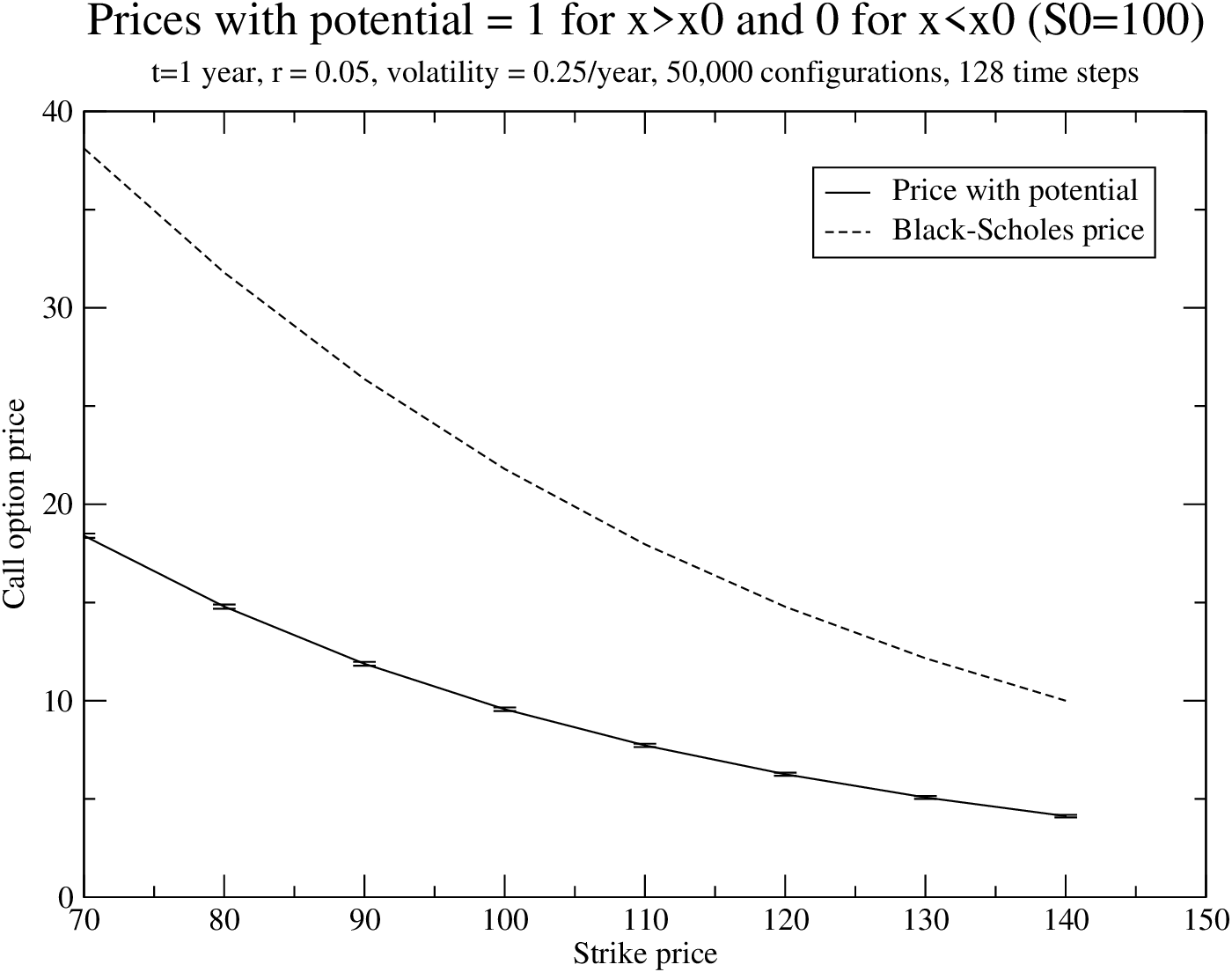, height=8cm}
  \caption{Plot of call option price against strike price for a purely
    discounting potential of form (\ref{eq:two}) and for the
    Black-Scholes case.}
  \label{fig:two}
\end{figure}

\begin{figure}[h]
  \centering
  \epsfig{file=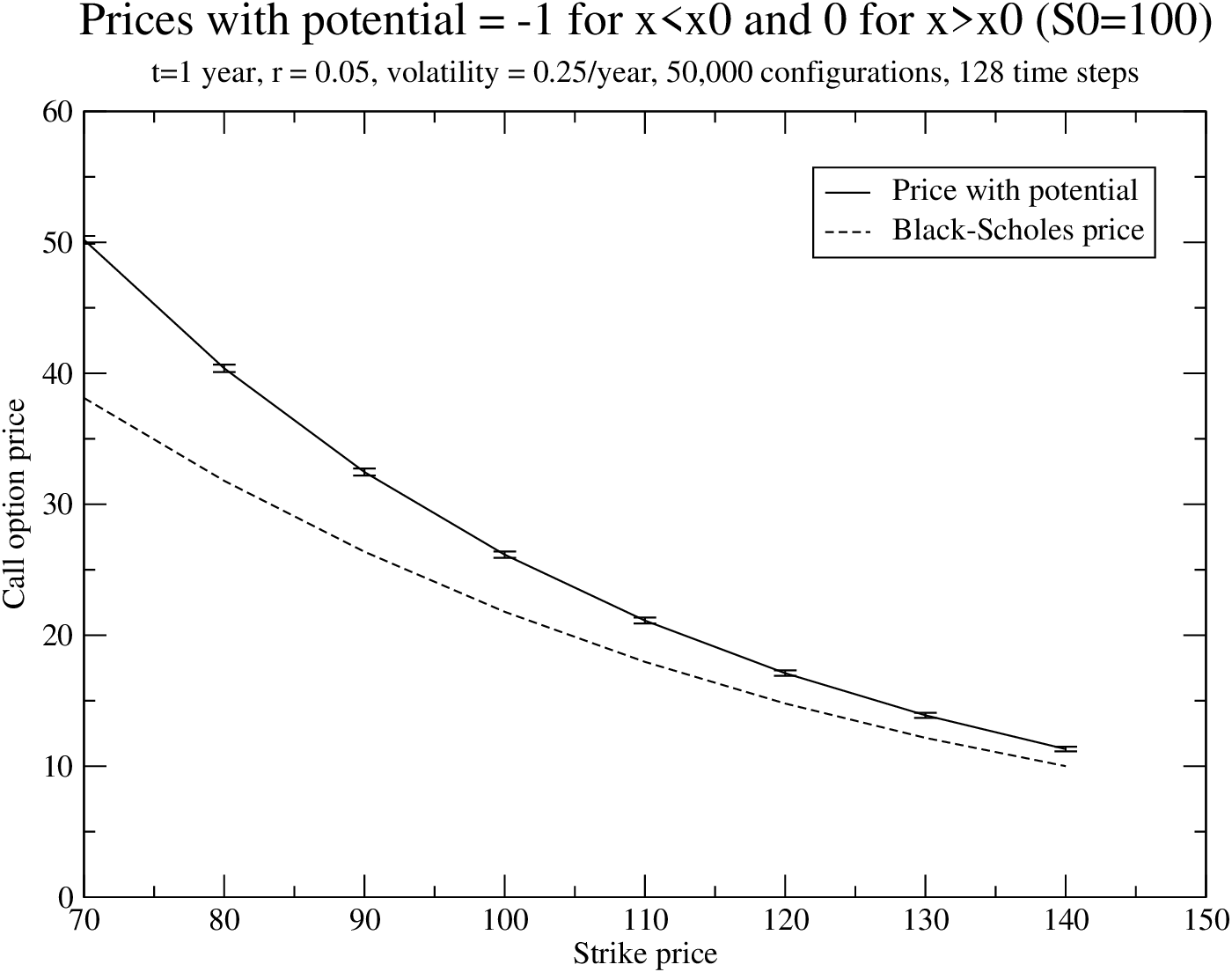, height=8cm}
  \caption{Plot of call option price against strike price for a purely
    discounting potential of form (\ref{eq:three}) and for the
    Black-Scholes case.}
  \label{fig:three}
\end{figure}

\begin{figure}[h]
  \centering
  \epsfig{file=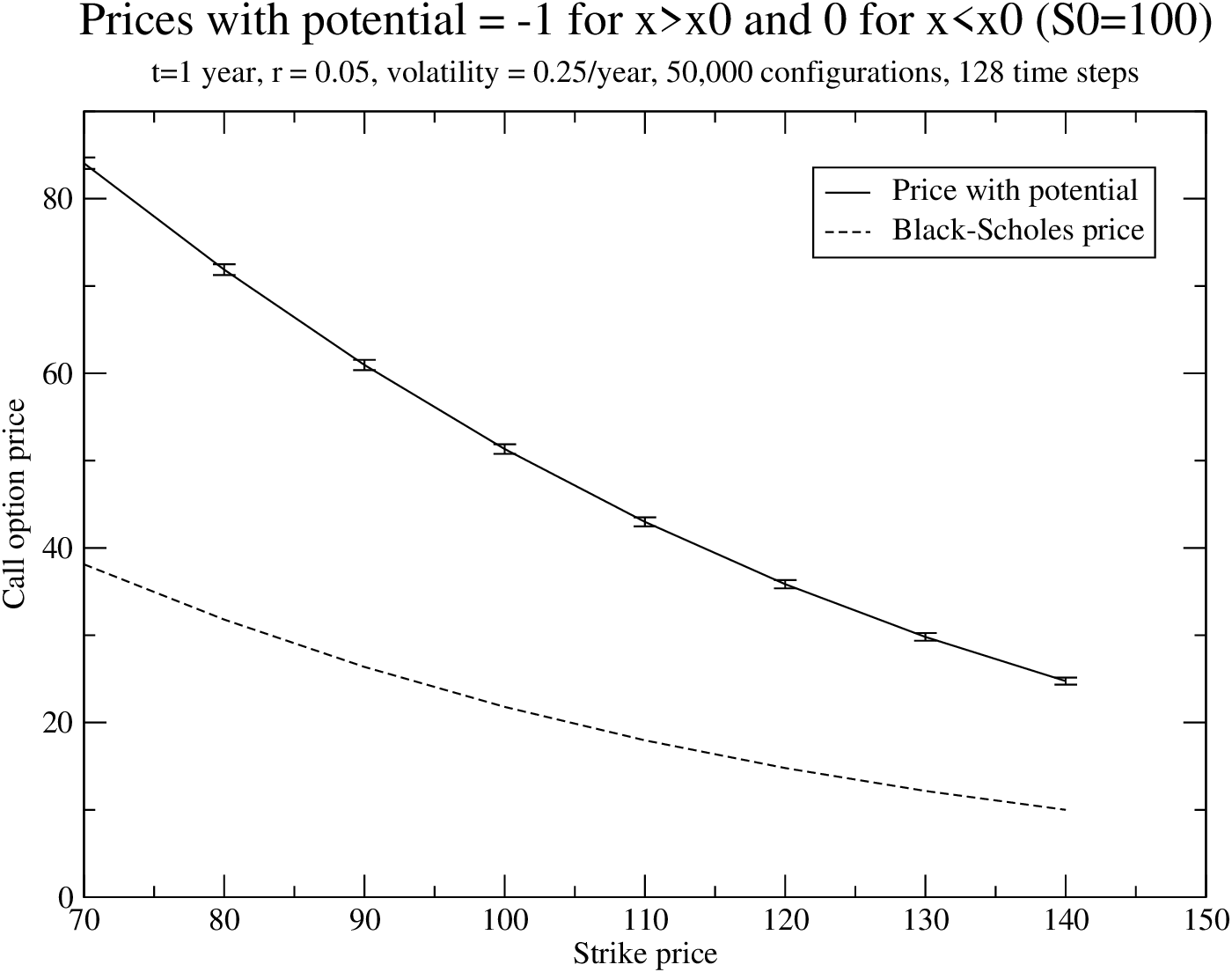, height=8cm}
  \caption{Plot of call option price against strike price for a purely
    discounting potential of form (\ref{eq:four}) and for the
    Black-Scholes case.}
  \label{fig:four}
\end{figure}

Figures \ref{fig:one}, \ref{fig:two}, \ref{fig:three} and
\ref{fig:four} refer to the potential as used in the first sense. The
interest rate was fixed at 5\%, the volatility $\sigma^2$ at
$0.25/$year and the time to expiry at one year. In figure
\ref{fig:one}, we show a plot of the call option price versus the
strike price with
\begin{equation}
  \label{eq:one}
  V(x) = 
  \begin{cases}
    1 & x<=x_0\\
    0 & x>x_0\\
  \end{cases}
\end{equation}
In figure \ref{fig:two}, the potential used is 
\begin{equation}
  \label{eq:two}
  V(x) = 
  \begin{cases}
    0 & x<=x_0\\
    1 & x>x_0\\
  \end{cases}
\end{equation}
while in figure \ref{fig:three}, it is 
\begin{equation}
  \label{eq:three}
  V(x) = 
  \begin{cases}
    -1 & x<=x_0\\
    0 & x>x_0\\
  \end{cases}
\end{equation}
and in figure \ref{fig:four}, it is
\begin{equation}
  \label{eq:four}
  V(x) = 
  \begin{cases}
    0 & x<=x_0\\
    -1 & x>x_0\\
  \end{cases}
\end{equation}

\begin{figure}[h]
  \centering
  \epsfig{file=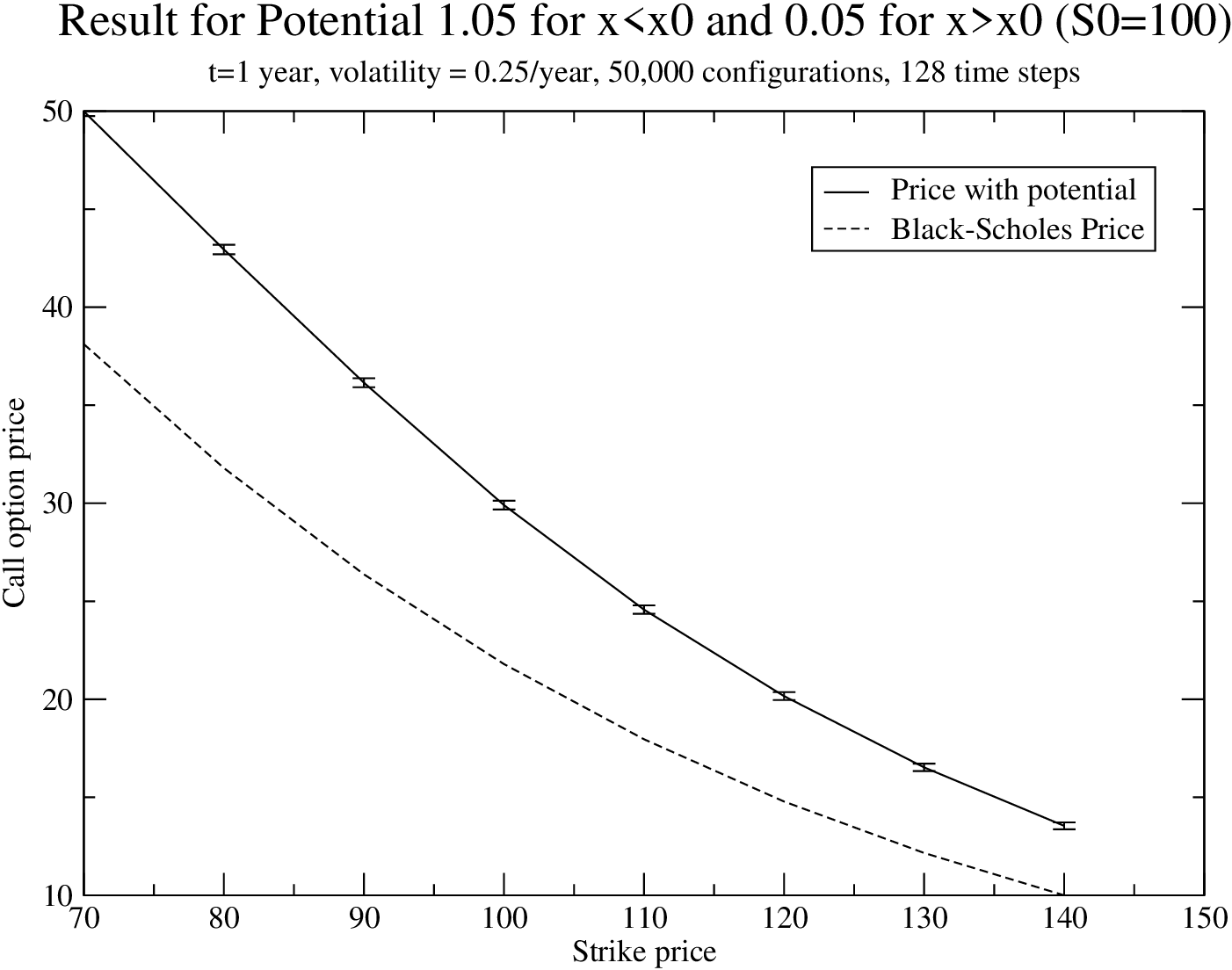, height=8cm}
  \caption{Plot of call option price against strike price for a
    potential of form (\ref{eq:five}) and for the Black-Scholes case.}
  \label{fig:five}
\end{figure}

\begin{figure}[h]
  \centering \epsfig{file=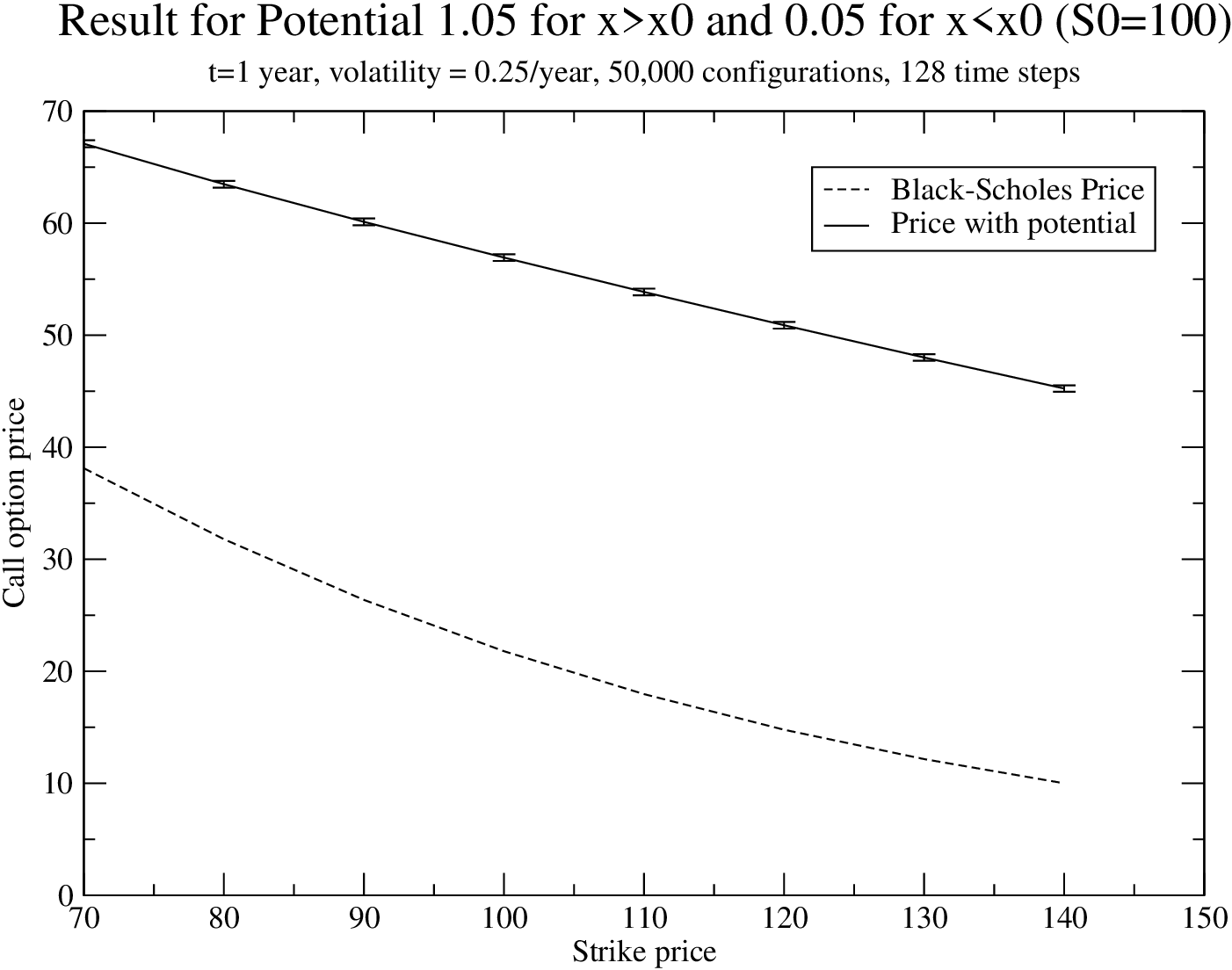, height=8cm}
  \caption{Plot of call option price against strike price for a
    potential of form (\ref{eq:six}) and for the Black-Scholes case.}
  \label{fig:six}
\end{figure}

\begin{figure}[h]
  \centering
  \epsfig{file=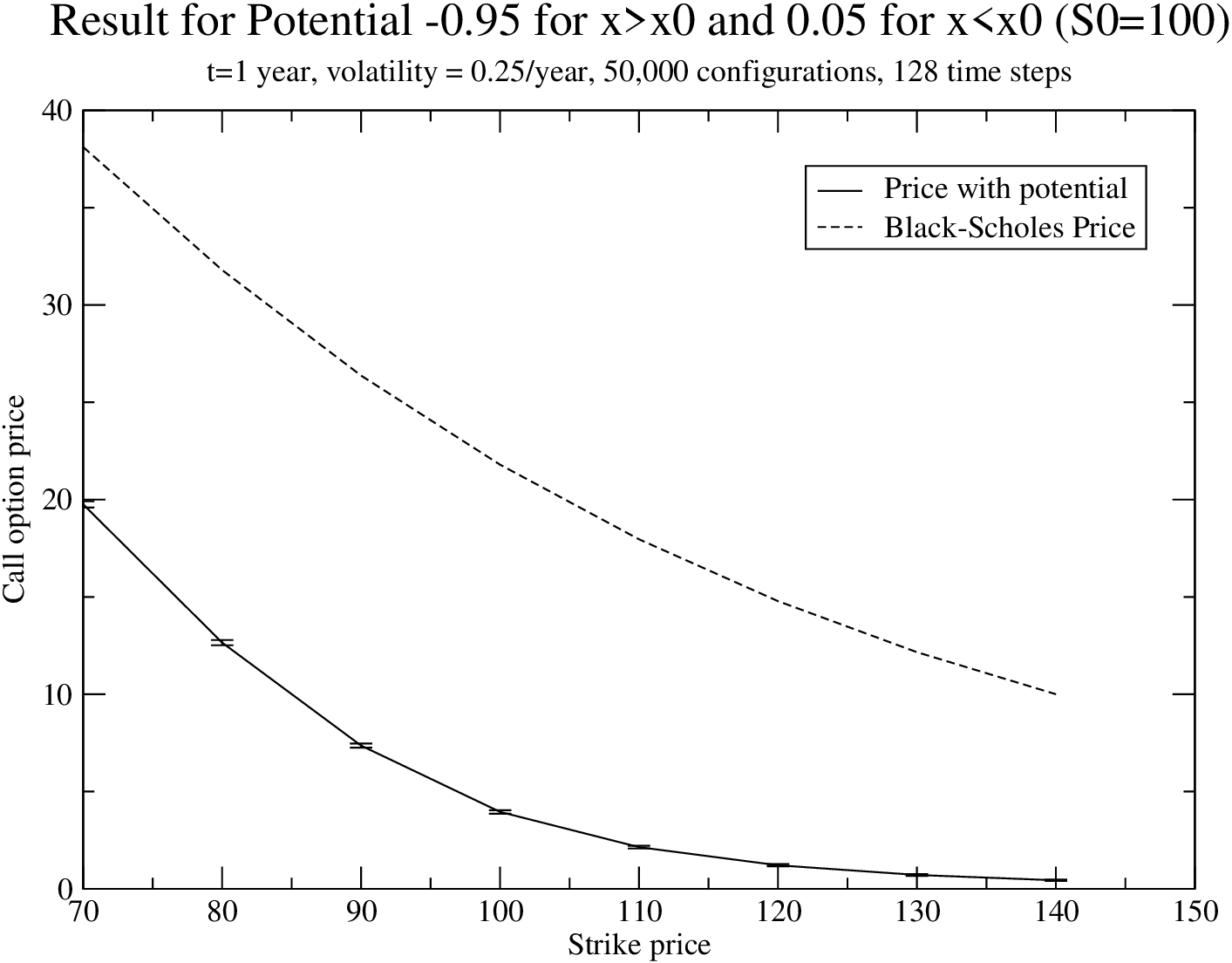, height=8cm}
  \caption{Plot of call option price against strike price for a
    potential of form (\ref{eq:seven}) and for the Black-Scholes
    case.}
  \label{fig:seven}
\end{figure}

\begin{figure}[h]
  \centering
  \epsfig{file=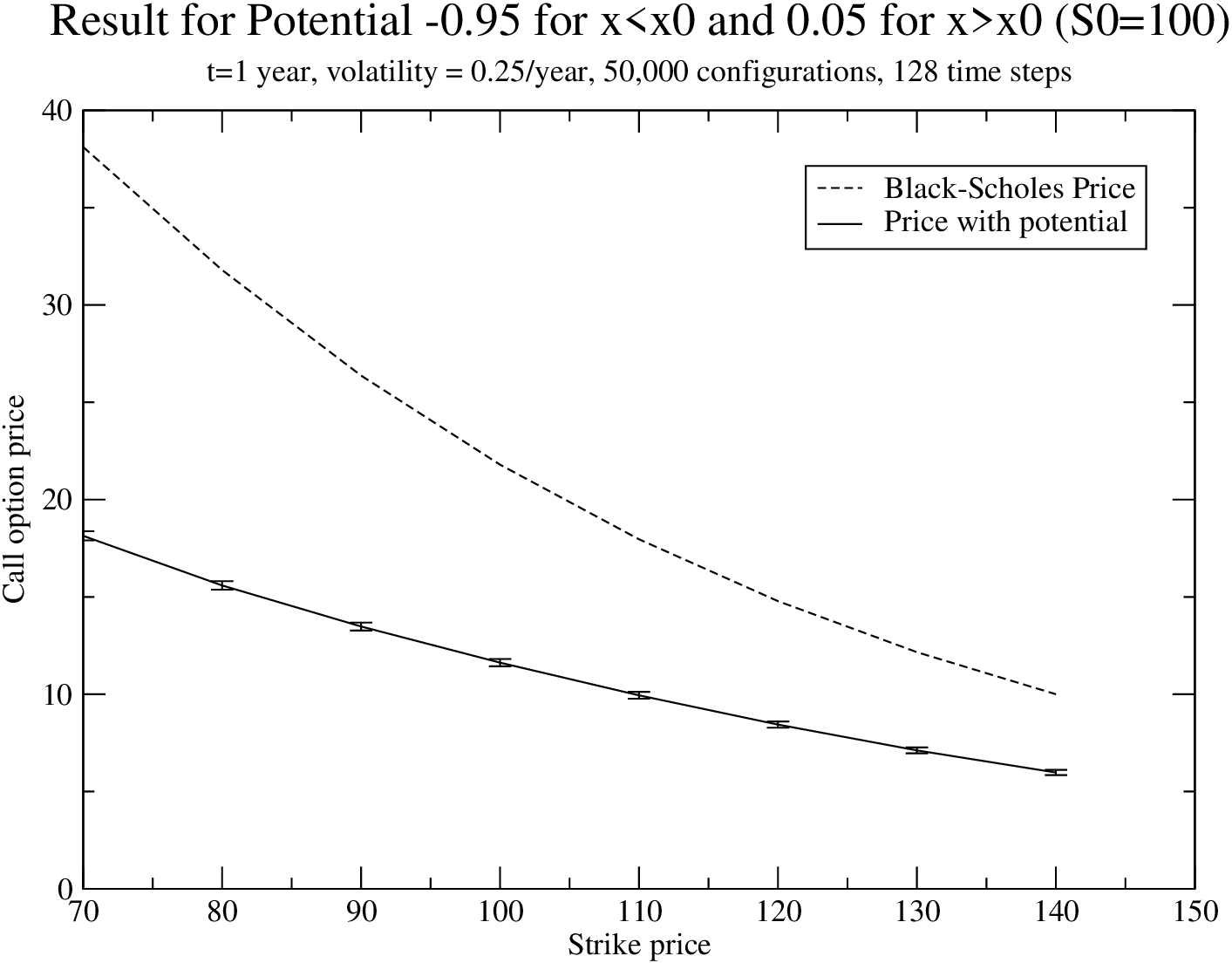, height=8cm}
  \caption{Plot of call option price against strike price for a
    potential of form (\ref{eq:eight}) and for the Black-Scholes
    case.}
  \label{fig:eight}
\end{figure}

Figures \ref{fig:five}, \ref{fig:six}, \ref{fig:seven} and
\ref{fig:eight} refer to the potential as used in the second sense.
The volatility at $0.25/$year and the time to expiry at one year. In
figure \ref{fig:five}, we show a plot of the call option price versus
the strike price with
\begin{equation}
  \label{eq:five}
  V(x) = 
  \begin{cases}
    1.05 & x<=x_0\\
    0.05 & x>x_0\\
  \end{cases}
\end{equation}
In figure \ref{fig:six}, the potential used is 
\begin{equation}
  \label{eq:six}
  V(x) = 
  \begin{cases}
    0.05 & x<=x_0\\
    1.05 & x>x_0\\
  \end{cases}
\end{equation}
while in figure \ref{fig:seven}, it is 
\begin{equation}
  \label{eq:seven}
  V(x) = 
  \begin{cases}
    -0.95 & x<=x_0\\
    0.05 & x>x_0\\
  \end{cases}
\end{equation}
and in figure \ref{fig:eight}, it is
\begin{equation}
  \label{eq:eight}
  V(x) = 
  \begin{cases}
    0.05 & x<=x_0\\
    -0.95 & x>x_0\\
  \end{cases}
\end{equation}
The last two cases are unrealistic in practice as interest rates can
never go negative.
\section{Conclusions}
The path integral formulation of financial instruments, as shown in this 
work, is a promising approach to the pricing of derivative products 
 which shows a remarkable flexibility. We have presented several applications 
of the method and have provided a general strategy -based on the use of 
a potential in the modeling of barrier options- to analyze these instruments. 

Specifically, we have compared Langevin simulations and Monte Carlo simulations and shown 
that a discount on the price of these options has indeed a rather simple 
interpretation in terms of paths of the underlying security. 
We believe that these strategies will turn out to be very effective 
for the simulation of complex portfolios, as well as for the inclusion of 
constraints in the evolution of these derivatives.

\centerline{\bf Acknowledgements}
C.C. thanks the University Scholars Program at the 
National University of Singapore for partial financial support, WBS number R-377-000-016-112 and Rajesh Parwani for discussions and hospitality. 
He thanks the High Energy Theory Group at the Univ. of Crete and in particular 
E. Kiritsis and T. Tomaras for discussions and hospitality while completing 
this work.

\section{Appendix A}
In this appendix we discuss some technical issues regarding the 
spectrum of the eigenvalue equation for the pricing of asian options. 

The general structure of the stationary Schrodinger equation for an Asian Option is of the form 
\beqa
 \label{eq:hbs}
&&  \left(\hat {H} - E\right)\psi(x)  \nonumber \\
&& = -\frac{\sigma^2}{2} \frac{\partial}{\partial x^2} +
  (\frac{\sigma^2}{2} - r) \frac{\partial}{\partial x} + \left( 
i K e^x -E\right)\psi =0 
\eeqa
which we rewrite in the form 
\beq
\left( \frac{\partial}{\partial x^2} -
  2 \alpha\frac{\partial}{\partial x} -\frac{2}{\sigma^2} \left( 
i K e^x -E\right)\right)\psi =0 
\eeq
with $\alpha = {(\sigma^2/2-r)}/{\sigma^2}$, as defined above. 
The velocity dependent term is eliminated as in the usual Black-Scholes model 
by factorizing an overall $e^{\alpha x}$ term in the eigenfunctions. 

The eigenfunctions for the Asian Option can be expressed in terms of modified Bessel functions of the 
first kind $I_\nu(z)$ 
and $I_{-\nu}(z)$, of complex argument $z$. Specifically, we introduce the 
index function 
\beq
\nu(E)\equiv \frac{\sqrt{\alpha^2 - 2 E}}{\sigma}
\eeq
then the solution is of the form 
\beqa
\psi_E(x)&= & 
C_1 e^{ \alpha x -i \frac{\pi}{4} \nu(E)} \Gamma\left(1 - 2 \nu(E )\right)I_{-2 \nu(E)}(z) + 
C_2 e^{ \alpha x + i \frac{\pi}{4} \nu(E)} \Gamma\left( 1 + 2 \nu(E)\right) I_{2 \nu(E)}(z), \nonumber \\ 
&& z=|z| e^{i \pi/4},  |z|= 2 \frac{\sqrt{2 K}}{\sigma} e^{x/2} \\
\eeqa
with $C_1$ and $C_2$  arbitrary integration constant. 
Notice that the eigenfunctions have a branch cut from $-\infty$ to $0$.


\end{document}